\documentstyle[twocolumn,epsfig]{mn}
\begin{document}

\title{The influence of winds on the time-dependent behaviour  of self-gravitating accretion discs}

\author[M. Shadmehri]{Mohsen Shadmehri\thanks{E-mail:
mohsen.shadmehri@dcu.ie (MS); }\\
Department of Mathematical Physics, National University Ireland Maynooth, Ireland\\
School of Mathematical Sciences, Dublin City University, Glasnevin, Dublin 9, Ireland}

\maketitle

\date{Received ______________ / Accepted _________________ }

\begin{abstract}

We study effects of winds on the time evolution of isothermal,  self-gravitating accretion discs by adopting a radius dependent mass loss rate because of the existence of the wind. Our similarity and semi-analytical solution describes time evolution of the system in the slow accretion limit. The disc structure is distinct in the inner and outer parts, irrespective of the existence of the wind. We show that existence of wind will lead to a reduction of the surface density in the inner and outer parts of the disc in comparison  to a no-wind solution. Also, the radial velocity significantly increases in the outer part of the disc, however, the accretion rate decreases due to the reduced surface density in comparison to the no-wind solution.  In the inner part of the disc, mass loss due to the wind is negligible according to our solution. But the radial size of this no-wind inner region becomes smaller for stronger winds.

\end{abstract}

\section{Introduction}
\label{sec:1}
The existence of outflow or wind in many accreting systems  is supported by strong observational evidences. Outflows from Active Galactic Nuclei (AGN) are much more common than previously thought: the overal fraction of AGNs with outflows is fairly constant, approximately $60\%$, over many order of magnitude in luminosity (Ganguly \& Brotherton  2008). These mass-loss mechanisms are also observed in microquasars, Young Stellar Object (YSO) and even from brown dwarfs (Ferrari 1998; Bally, Reipurth \& Davis 2007; Whelan et al 2005). About 30$\%$ of T Tauri stars present bipolar ejection and this percentage increases to $100\%$ for class $0$ objects, the earliest stage of star formation.  It is now widely accepted that winds or outflows have their origin in accretion flows (e.g., Blandford \& Payne 1982). They may provide an additional important sink of mass, angular momentum and energy which their dynamical influence to the accretion flow can not be neglected. However, it remains unclear as how some  part of the accreting gas can be transferred into winds or outflows.

Nevertheless, deviations from Keplerian rotation in some AGNs and the flat infrared spectrum of some T Tauri stars can both be described by self-gravitating discs. Vertical structure of an accretion disc under the influence of self-gravity was studied by Paczy$\acute{\rm n}$ski (1978). Some authors studied the effects related to self-gravity of the disc in the radial direction (e.g., Bodo \& Curir 1992). New class of self-gravitating discs has also been proposed, in which the energy equation is replaced by a self-regulation condition (Bertin \& Lodato 1999; Lodato \& Bertin 2001). Mineshige \& Umemura (1997) extended the previous steady state solutions to the time-dependent case while the effect of the self-gravity of the disc was taken into account. They used an isothermal equation of state, and so their solutions describe a viscous accretion disc in the slow accretion limit. Also, Tsuribe (1999) studied the self-similar collapse of an isothermal viscous accretion disc.

For simplicity, outflow is neglected in most of the steady-state or time-dependent theoretical models of the self-gravitating discs. However, some authors studied the effect of wind or outflow on the radial structure of non-self-gravitating accretion discs (e.g., Knigge 1999). Recently, Combet \& Ferreira (2008) studied the structure of YSO accretion discs in an approach that takes into account the presence of the protostellar jets. They showed that discs with jet  presents structure different from the standard accretion disc  because of the influences of jets on the radial structure of the disc. Ruden (2004) studied the physics of protoplanetary disc evolution in the presence of a photoevaporative wind.

In this paper, we generalize time-dependent solutions of Mineshige \& Umemura (1997) for a self-gravitating accretion disc to include winds or outflows. Our parametric model describes mass and angular momentum loss by wind in an isothermal, self-gravitating disc, yet applicable to many types of dynamical disc-wind models. Basic equations are presented in the next section. Properties of our semi-analytical are discussed in section 3. A summary of implications of the results are given in section 4.

\section{General Formulation}
\label{sec:1}
We consider an accretion disc that is axisymmetric and geometrically thin, i.e. $H/R < 1$. Here $R$ and $H$ are, respectively, the disk radius and the half-thickness. The disc is supposed to be turbulent and possesses an effective turbulent viscosity. In our model, a central object has not yet been formed and the radial component of the gravitational force is provided by the self-gravity of the disc. The continuity equations reads
\begin{equation}\label{eq:main1}
\frac{\partial \Sigma}{\partial t}+\frac{1}{r}\frac{\partial}{\partial r}(r\Sigma v_{\rm r})+\frac{1}{2\pi r}\frac{\partial \dot{M}_{\rm w}}{\partial r}=0,
\end{equation}
where $v_{\rm r}$ is the accretion velocity ($v_{\rm r}<0$) and $\Sigma = 2\rho H$ is the surface density at a cylindrical radius $r$. Also, $\rho$ is the midplane density of the disc and the mass loss rate by outflow/wind is represented by $\dot{M}_{\rm w}$. So,
\begin{equation}
\dot{M}_{\rm w}(R) = \int 4\pi R' \dot{m}_{\rm w} (R') dR',\label{eq:mdot}
\end{equation}
where $\dot{m}_{\rm w} (R)=\rho v_{\rm z}^{+}$ is mass loss rate per unit area from each disc face. Here, $v_{\rm z}^{+}$ is a mean vertical velocity at the disc surface, i.e., at the base of a wind.

The radial momentum equation is
\begin{equation}\label{eq:main2}
\frac{\partial v_{\rm r}}{\partial t} + v_{\rm r}\frac{\partial v_{\rm r}}{\partial r}=-\frac{c_{\rm s}^2}{\Sigma}\frac{\partial\Sigma}{\partial r}-\frac{GM_{\rm r}}{r^2}+\frac{v_{\varphi}^2}{r},
\end{equation}
where $v_{\varphi}$ is the rotational velocity. As in Mineshige \& Umemura (1997),
we adopt the monopole approximation for the radial gravitational force due to the self-gravity of the disc, which considerably simplify the
calculations and are not expected to introduce any significant errors as long as the surface density profile is steeper than $1/r$
(e.g., Li \& Shu 1997; Saigo \& Hanawa 1998; Tsuribe 1999; Krasnopolsky \& K\"{o}nigl 2002). Also, we assume that the disc is vertically self-gravitating and so, the half thickness of the disc, $H$, becomes $H=c_{\rm s}^{2}/ (2\pi G \Sigma)$.

Similarly, integration over $z$ of the azimuthal equation of motion gives (e.g., Knigge 1999)
\begin{displaymath}
\frac{\partial}{\partial t} (r v_{\varphi})+v_{\rm r}\frac{\partial}{\partial r} (rv_{\varphi})=\frac{1}{r\Sigma}\frac{\partial}{\partial r}(r^3 \nu \Sigma \frac{\partial\Omega}{\partial r})
\end{displaymath}
\begin{equation}\label{angularmain}
-\frac{(lr)^2\Omega}{2\pi r\Sigma}\frac{\partial \dot{M}_{\rm w}}{\partial r},
\end{equation}
where the last term of right hand side represents angular momentum carried by the outflowing material. Here, $l=0$ corresponds to a non-rotating wind and $l=1$ to outflowing material that carries away the specific angular momentum it had at the point of ejection and it should
be most appropriate for radiation-driven outflows (Knigge 1999). Centrifugally driven MHD winds are corresponding to $l>1$ as has been discussed in Knigge (1999). In this case, we have $l=R_{\rm A}/R$ where $R_{\rm A}$ is Alfven radius (e.g., Knigge 1999). 
 Also, $\nu$ is a kinematic viscosity coefficient and we assume $\nu = \alpha c_{\rm s} H = \alpha (H/r)c_{\rm s} r$, where $c_{\rm s}$ is the sound speed. As in Mineshige \& Umemura (1997), we assume $\alpha'=\alpha (H/r)$ is constant in space. Thus, in our model the viscosity coefficient is in proportion to the radius which has also been used by some other authors (e.g., Hartmann et al 1998).
We introduce similarity variable $x\equiv r/(c_{\rm s}t)$ and the physical quantities as
\begin{equation}
\Sigma (r, t)= \frac{c_{\rm s}}{2\pi G t} \sigma (x),
\end{equation}
\begin{equation}
v_{\rm r}(r, t)= c_{\rm s} u(x),
\end{equation}
\begin{equation}
v_{\varphi}(r, t) = c_{\rm s} v(x),
\end{equation}
\begin{equation}
j = xv(x),
\end{equation}
\begin{equation}
\dot{m}_{\rm w}=\frac{c_{\rm s}}{4\pi G t^2} \sigma(x) \Gamma(x).
\end{equation}
Thus, the accretion rate becomes $\dot{M}_{\rm acc}=-2\pi r\Sigma v_{\rm r}=(c_{\rm s}^{3}/G)\dot{m}$, where $\dot{m}=x(-u)\sigma$ is the non-dimensional similarity accretion rate.
 Now, we can write equations (\ref{eq:main1}), (\ref{eq:main2}) and (\ref{angularmain}) as
\begin{equation}
-\sigma - x\frac{d\sigma}{dx}+\frac{1}{x}\frac{d}{dx}(x\sigma u)+ \sigma \Gamma =0,\label{eq:con}
\end{equation}
\begin{equation}
\frac{2}{\sigma}\frac{d\sigma}{dx}+(u-x)\frac{du}{dx}-\sigma\frac{u-x}{x}-\frac{v^2}{x}=0,\label{eq:radial}
\end{equation}
\begin{equation}
(l^{2}\Gamma +1)j + (u-x)\frac{dj}{dx}=\alpha' \frac{1}{\sigma x}\frac{d}{dx}[\sigma x (-2j+x\frac{dj}{dx}) ].\label{eq:angular}
\end{equation}

We can solve the above differential equations numerically subject to appropriate asymptotic behaviors as boundary conditions. But we restrict to solution in slow accretion limit which implies $v\gg 1$, $\sigma \gg 1$ and $|u|\ll 1$. Then, equation (\ref{eq:radial}) gives $v=\sigma^{1/2}(x-u)^{1/2}$ or
\begin{equation}
j=\sigma^{1/2} x (x-u)^{1/2}.\label{eq:j}
\end{equation}
On the other hand, equation (\ref{eq:con}) can be rewritten as
\begin{equation}
\frac{d\ln \sigma}{d\ln x}=\frac{1}{x-u}\frac{du}{d\ln x}-1 - \frac{x\Gamma}{u-x}. \label{eq:sigx}
\end{equation}

Also, after mathematical manipulation, from equation (\ref{eq:j}) we have
\begin{equation}
\frac{d\ln j}{d\ln x}=1+\frac{1}{2}\frac{1}{x-u}(x-\frac{du}{d\ln x})+\frac{1}{2}\frac{d\ln \sigma}{d\ln x}. \label{eq:jx}
\end{equation}
Substituting equation (\ref{eq:sigx}) into equation (\ref{eq:jx}), we obtain
\begin{equation}
\frac{d\ln j}{d\ln x}=\frac{2x-u+x\Gamma}{2(x-u)}. \label{eq:jxx}
\end{equation}

Using equation (\ref{eq:jxx}) the similarity angular momentum equation (\ref{eq:angular}) is written as
\begin{equation}
\frac{u}{x}+(2l^{2}-1) \Gamma = \alpha' \frac{1}{\sigma x j}\frac{d}{dx}[\sigma x j (\frac{3u-2x+x \Gamma}{x-u})].\label{eq:f}
\end{equation}
Equations (\ref{eq:sigx}) and (\ref{eq:jxx}) give
\begin{equation}
\frac{d\ln(\sigma x j)}{d\ln x}=\frac{1}{x-u}\frac{du}{d\ln x}+\frac{2x-u+3x\Gamma}{2(x-u)}.
\end{equation}
The above relation helps us to simplify equation (\ref{eq:f}) as
\begin{equation}
\frac{du}{dx}=-\frac{Ax^2+Bux+3u^2}{2(x-3u-2x\Gamma)x}-\frac{(x-u)^2 [u+(2l^2-1)x\Gamma)]}{\alpha' (x-3u-2x\Gamma)x},\label{eq:final}
\end{equation}
where
\begin{displaymath}
A=4+6\Gamma-2\frac{d}{dx}(x\Gamma)-3\Gamma^2,
\end{displaymath}
\begin{displaymath}
B=-2[3+4\Gamma-\frac{d}{dx}(x\Gamma)].
\end{displaymath}
First order differential equation (\ref{eq:final}) is the main equation of our analysis which can be solved numerically. Having profile of $u(x)$ from equation (\ref{eq:final}), the similarity surface density variable is calculated using equation (\ref{eq:sigx}) numerically. Clearly, the effect of wind or outflows appears by the term $\Gamma$. If we set this parameter equal to zero, equation (\ref{eq:final}) reduces to equation (18) of Mineshige \& Umemura (1997) which describes no-wind solution.

Behavior of the solutions with winds highly depends on the profile of the mean vertical velocity at the disc surface, i.e. $\Gamma$.
However, asymptotic behavior of the solutions near to the central part of the disc, i.e. $x\rightarrow 0$, is similar to a case without
wind/outflow according to equations (\ref{eq:sigx}) and  (\ref{eq:final}). When $x$ tends to zero, we have
\begin{equation}
|u| \propto x, \sigma \propto x^{-5/3}, v\propto x^{-1/3}.
\end{equation}
Appropriate boundary condition at $x=0$ is determined using the above asymptotic behavior. For starting the integration of equation (\ref{eq:final}), we assume $u=0$ at $x=0$. But in order to determine surface density profile, we can integrate equation (\ref{eq:sigx}) from outer boundary of the disc, i.e. $x=1$, towards the center for a given $\sigma(x=1)=\sigma_{0}$. But we consider a series of models where the accretion rate at the outer edge (i.e., the inflow rate from the parent cloud) is kept constant, which appears a natural requirement. Thus, our boundary conditions are $u=0$ at $x=0$, and, $\dot{m}=\dot{m}_{0}$ at $x=1$. So, there is another input parameter for our model, i.e. $\dot{m}_{0}$. For the mass loss profile, we consider a simple power-law form for $\Gamma$ as $\Gamma =\Gamma_{0} x^{s}$. We find two different regimes according to our similarity solution: inner no-wind part and an outer region with the surface density profile in proportion to $x^{-1}$. Thus, our prescription for  $v_{\rm z}^{+}/c_{\rm s}=\Gamma / \sigma$ gives us a power law mass loss rate for the wind as $\dot{M}_{\rm w} \propto x^{s+1}$. In steady state case, this prescription for mass loss by wind has been used widely by many authors (e.g.,  Quataert \& Narayan 1999; Beckert 2000; Turolla \& Dullemond 2000; Misra \& Taam 2001; Fukue 2004).

\begin{figure*}
%\vspace{-50pt}
\epsfig{figure=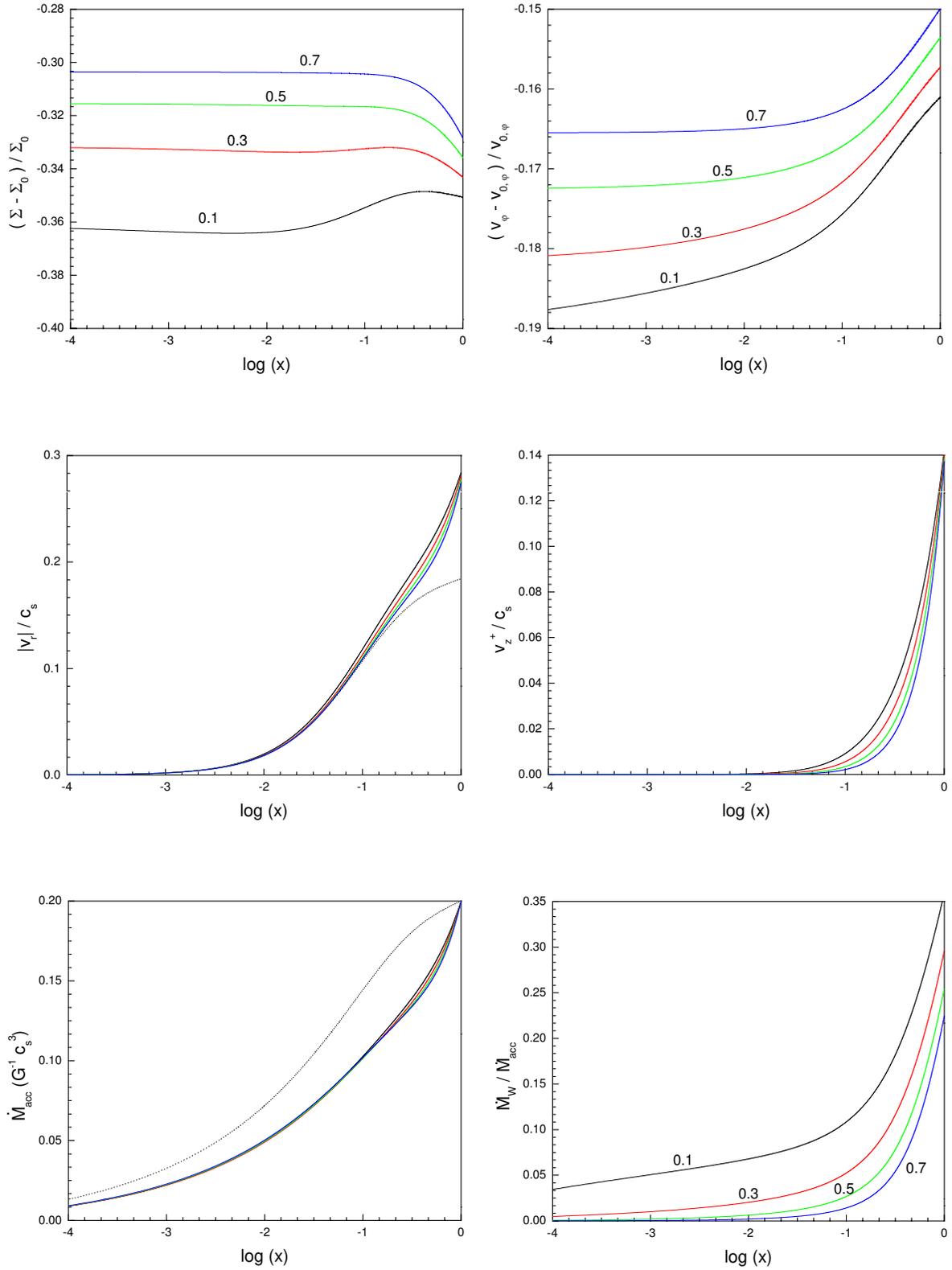,angle=0,scale=0.85}
\caption{The profiles of the physical variables for $\alpha' = 0.1$, $\dot{m}_{0}=0.2$ and $s=0.1$, $0.3$, $0.5$, $0.7$  with $\Gamma_{0}=0.1$ and $l=1$ (i.e, rotating wind). Surface density and the rotational velocity for no-wind solution are represented by $\Sigma_{0}$ and $v_{0, \varphi}$. Each curve is labeled by corresponding $s$. No-wind solution is shown by dotted curves. }
\label{fig:f1}
\end{figure*}

\begin{figure*}
%\vspace{-10pt}
\epsfig{figure=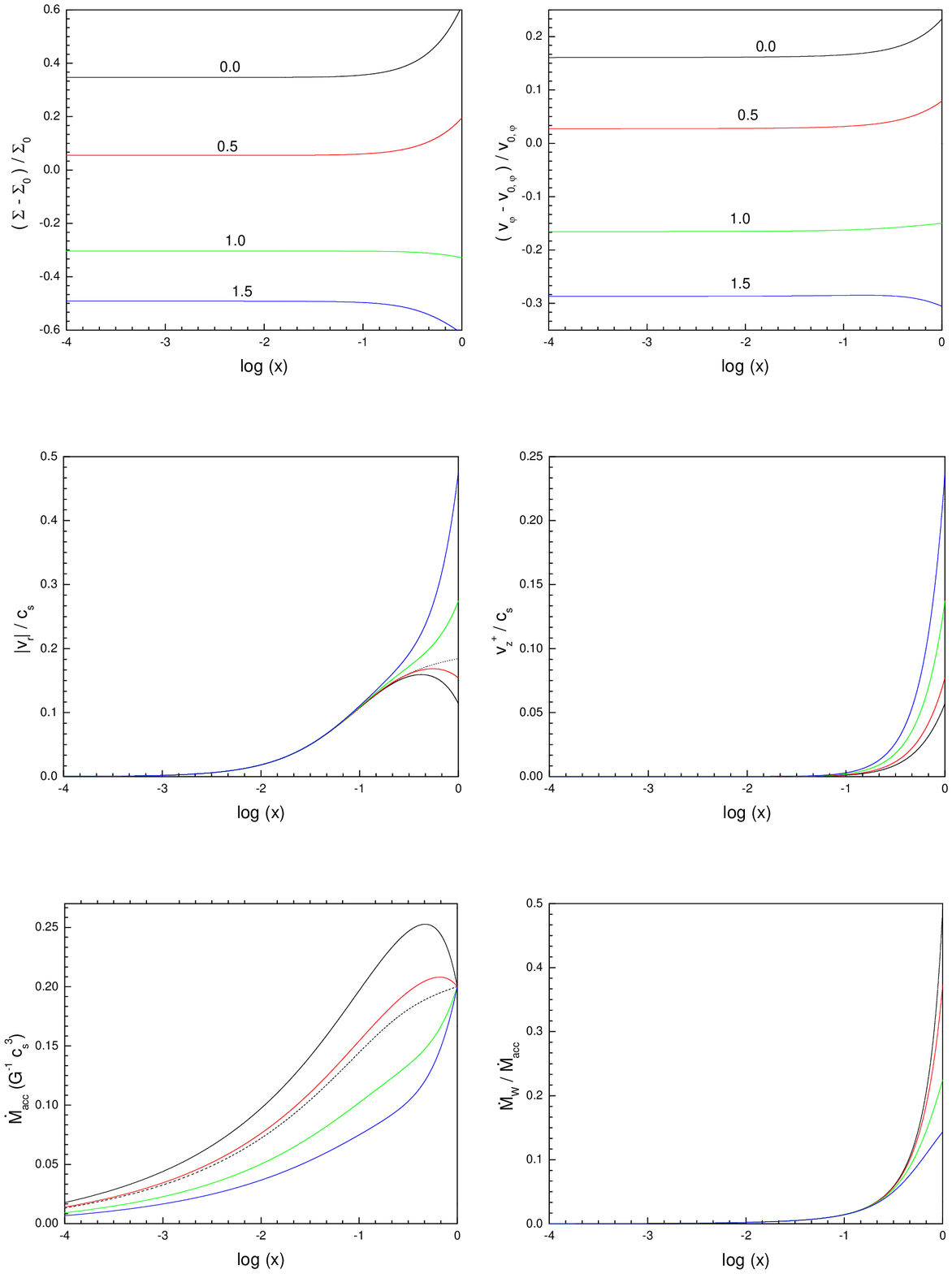,angle=0,scale=0.85}
\caption{The profiles of the physical variables for $\alpha' = 0.1$, $\dot{m}_{0}=0.2$ and $l=0.0$, $0.5$, $1.0$, $1.5$  with $\Gamma_{0}=0.1$ and $s=0.7$. Surface density and the rotational velocity for no-wind solution are represented by $\Sigma_{0}$ and $v_{0, \varphi}$. Each curve is labeled by corresponding $l$. No-wind solution is shown by dotted curves.}
\label{fig:f2}
\end{figure*}

\begin{figure*}
%\vspace{-10pt}
\epsfig{figure=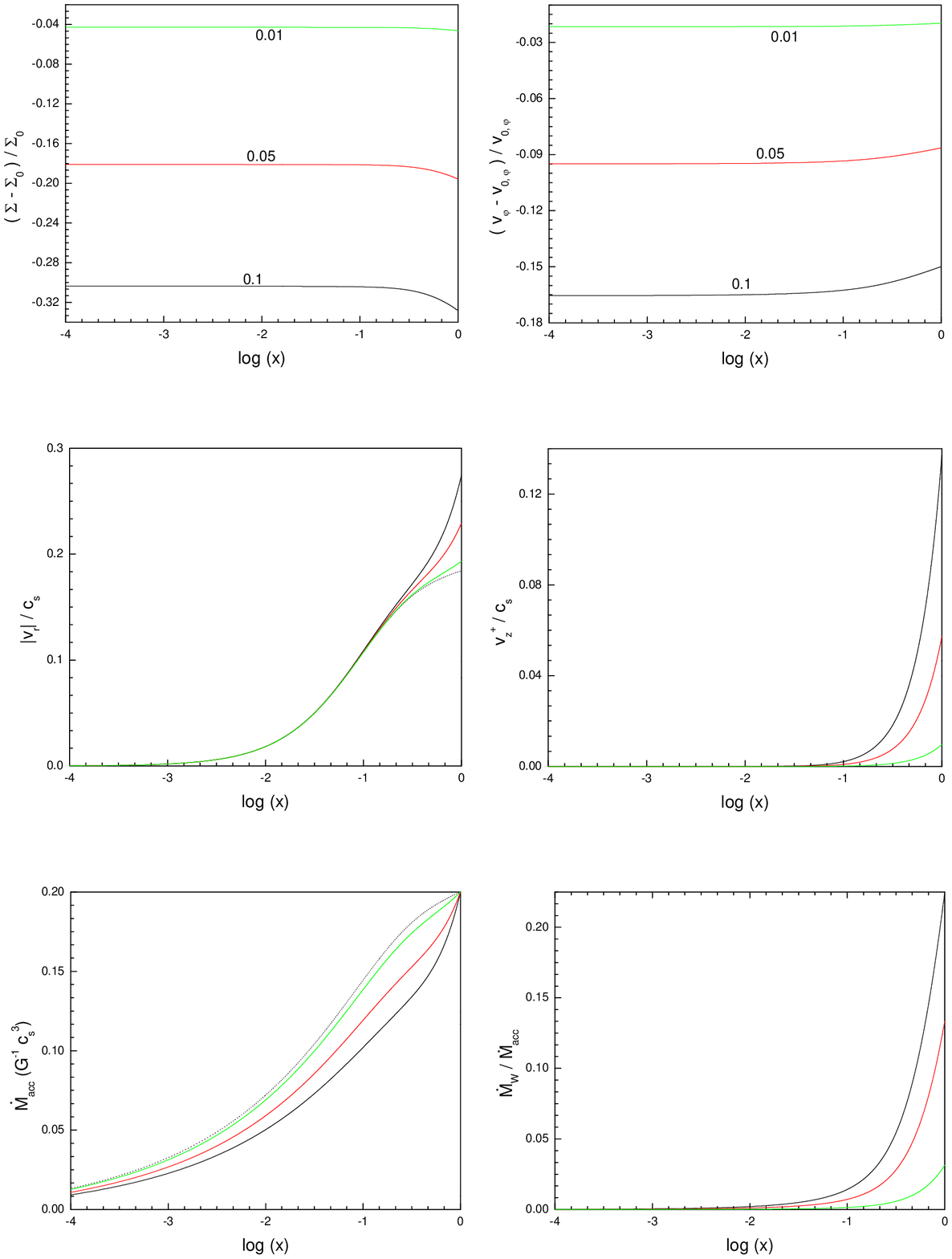,angle=0,scale=0.85}
\caption{The profiles of the physical variables for $\alpha' = 0.1$, $\dot{m}_{0}=0.2$ and $\Gamma_{0}=0.1$, $0.05$, $0.01$  with $l=1.0$ and $s=0.7$. Surface density and the rotational velocity for no-wind solution are represented by $\Sigma_{0}$ and $v_{0, \varphi}$. Each curve is labeled by corresponding $\Gamma_{0}$. No-wind solution is shown by dotted curves.}
\label{fig:f3}
\end{figure*}

\section{analysis}
Note that we are doing a purely parametric approach to take into account effects of wind on a self-gravitating accretion disc. So, this does not ensure that a self-consistent solution exists for any given set of the specified parameters. For solving the equations, we check out that  solutions do not violate two important constraints: (a) the mass loss rate by the wind must be less than the accretion rate,  (b) the slow accretion limit is satisfied. We restrict our study to positive values for the exponent $s$, because  for $s<0$ we found qualitatively similar results. Note that in our model we have $\dot{M}_{\rm w} \propto x^{s+1}$ and $x\leq 1$ which imply a stronger  outflow for a smaller positive index $s$ .  Our parameterized approach is nevertheless useful, because it illustrates the possible effects of winds on the time-dependent structure of a self-gravitating disc.

Figure \ref{fig:f1} shows the change in the profiles of the physical variables with the mass loss power-law index $s$. Each curve is labeled by corresponding index $s$. Also, we adopt $\alpha'=0.1$, $\dot{m}_{0}=0.2$, $\Gamma_{0}=0.1$ and $l=1$ (i.e., rotating wind). The surface density and the rotational velocity for the no-wind solution are represented by $\Sigma_{0}$ and $v_{0,\varphi}$. We find that the structure is represented by an inner region with a density profile in proportion to $x^{-5/3}$ and an outer part with density profile proportional to $x^{-1}$, irrespective of the existence of wind or outflow. The transition radius at $x_{\rm tr}\sim \alpha'$ separates the inner and outer parts. However, the surface density decreases at all parts of the disc because of the wind. In order to make an easier comparison, the ratio $(\Sigma - \Sigma_{0})/\Sigma_{0}$ versus the similarity variable $x$ is shown in Figure \ref{fig:f1} (top, left). The surface density reduction  is more significant  for smaller values of the exponent $s$. As the wind becomes stronger and more mass is extracted from the disc, reduction to the surface density is more significant.

Profile of the rotational velocity versus the similarity variable is also shown in Figure \ref{fig:f1} (top, right). Generally, solutions with winds will rotate slower than those without winds. So, the viscous dissipation per unit mass in the flow is expected to be smaller in the presence of a wind. The reduced rotational velocity of the disc is sensitive to the variations of the exponent $s$. Also, profile of the radial velocity in Figure \ref{fig:f1} (middle, left) shows significant deviations from no-wind solution because of the presence of winds. The radial velocity is approximately uniform in  the outer parts of a disc without winds. But when the wind carries away the angular momentum appropriate to the radius from which it is launched, the remaining gas in the outer parts of the disc has much larger radial velocity in comparison to the no-wind solution. As  wind becomes stronger, deviations of the radial velocity from the no-wind solution is occurring over a larger region of the disc. Velocity of the wind at the surface of the disc is shown in Figure \ref{fig:f1} (middle, right). We can simply show that $v_{z}^{+}/c_{\rm s} \propto x^{s+1}$ at the outer part of the disc. When the exponent $s$ increases, velocity of the wind at the surface of the disc decreases.

Profile of the accretion rate, $\dot{M}_{\rm acc}$, is shown in Figure \ref{fig:f1} (bottom, left). Accretion rate for the no-wind solution is represented by the dotted lines. We can see that the accretion rate decreases at all parts of the disc due to the existence of the wind. However, the accretion rate is not very sensitive to the variations of the exponent $s$.   Ratio of the mass loss rate due to the wind to the accretion rate, i.e. $\dot{M}_{\rm w}/\dot{M}_{\rm acc}$, is represented in Figure \ref{fig:f1} (bottom, right). In the inner part of the disc, the mass loss by wind is negligible except for a strong wind corresponding to $s=0.1$. We adopted the input parameters so that the ratio is less than one at all radii of the disc. But most of the mass loss by wind is occurring at large radii, i.e. outer part of the disc. For a stronger wind, a larger fraction of the mass carries away by the wind.

We explore possible effects of extraction of angular momentum in Figure \ref{fig:f2} by changing the parameter $l$. In this figure, we assume $\alpha' = 0.1$, $\dot{m}_{0}=0.2$ and $l=0.0$, $0.5$, $1.0$, $1.5$  with $\Gamma_{0}=0.1$ and $s=0.7$. Obviously, angular momentum is not extracted by the wind when we have $l=0$. Actually, this case corresponds to a non-rotating wind and the disc losses only mass because of the wind. However, we found that for $l^{2}<1/2$ the mass of the disc increases in the presence of the winds that is obviously unphysical. It is partly because of the limitations of similarity method that there is not a self-consistent solution for any given set of the input parameters, and  more importantly, our model is valid in the slow accretion limit (i.e., $v \gg 1$), and so it is very unlikely to accept that winds are lunched  without extracting a certain amount of angular momentum of the disc.  Although solutions with $l=0$ and $0.5$ are represented in Figure \ref{fig:f2} for making a comparison, we think, these solutions are not physically acceptable. Surface density profile is shown in Figure \ref{fig:f2} (top, left). Again, we see reduction to the surface density because of the wind. Also, rotational velocity of the disc decreases when there is significant angular momentum loss by the wind (top, right). Although the radial velocity of the inner disc does not change because of the wind, existence of the a rotating wind enhances the radial velocity at the outer part of the disc in comparison to a no-wind solution (middle, left). Typical behavior of velocity $v_{\rm z}^{+}$ is also sensitive to the amount of the extracted angular momentum (middle, right). Profile of the accretion rate (bottom, left) shows that the accretion rate decreases due to the existence of a rotating wind. However, as more angular momentum is carried away by the wind (i.e. larger $l$), not only the surface density and the rotational velocity are reduced  at all regions of the disc, but  the radial velocity is increases significantly at the outer part of the disc.

Another input parameter of our model is $\Gamma_0$ that its possible effects are explored in Figure \ref{fig:f3}. We assume that $\alpha' = 0.1$, $\dot{m}_{0}=0.2$ and $\Gamma_{0}=0.1$, $0.05$, $0.01$  with $l=1.0$ and $s=0.7$. Again, the surface density and the rotational and radial velocities are significantly decreasing with $\Gamma_0$. The velocity of the wind at the surface of the disc is highly affected by the parameter $\Gamma_0$ (middle, right). Thus, the accretion rate and the mass loss by the wind are decreased with the parameter $\Gamma_0$.

\begin{figure}
%\vspace{-50pt}
\epsfig{figure=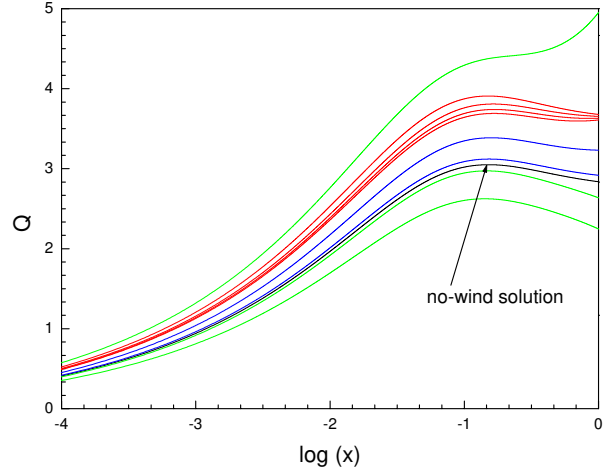,angle=0,scale=0.55}
\caption{The profile of Toomre parameter versus similarity variable for the solutions which are represented in Figures \ref{fig:f1}, \ref{fig:f2} and \ref{fig:f3}. It is assumed that $\dot{m}_{0}=0.2$. {\it Red curves} are corresponding to the solutions for $\alpha' = 0.1$ and $s=0.1$, $0.3$, $0.5$, $0.7$  with $\Gamma_{0}=0.1$ and $l=1$. Toomre parameter for the solutions  with $\alpha' = 0.1$ and $l=0.0$, $0.5$, $1.5$  with $\Gamma_{0}=0.1$ and $s=0.7$ are represented by {\it green curves}. Also, {\it blue curves} are showing Toomre parameter when  $\alpha' = 0.1$ and $\Gamma_{0}=0.01$, $0.05$  with $l=0.1$ and $s=0.7$.}
\label{fig:f4}
\end{figure}

The effect of the disc self-gravity in this paper is limited to provide the radial gravitational field to keep the disc in centrifugal equilibrium. On the other hand, it is well known that self-gravitating discs can be unstable if they are too cold. We can study gravitational stability of the solutions using the Toomre criteria (Toomre 1964),
\begin{equation}
Q=\frac{c_{\rm s} \kappa}{\pi G \Sigma},
\end{equation}
where $\kappa = \sqrt{2(v_{\varphi}/r^{2}) d(rv_{\varphi})/dr}$ is the epicyclic frequency. We calculate Toomre parameter $Q$ using our solutions and the disc is gravitationally stable if $Q>1$. We can rewrite Toomre parameter in terms of the similarity quantities as
\begin{equation}
Q = \frac{2\sqrt{2}}{x\sigma} \sqrt{\frac{j}{x}\frac{dj}{dx}}.\label{eq:Toomre}
\end{equation}

Figure \ref{fig:f4} shows profile of  Toomre parameter for the solutions which are presented in Figures \ref{fig:f1}, \ref{fig:f2} and \ref{fig:f3}. Here, we assumed that the nondimensional similarity accretion rate at the outer boundary is $0.2$, i.e. $\dot{m}_{0}=0.2$. To make an easier comparison, Toomre parameter for a case without wind/outflow is represented in Figure \ref{fig:f4}. This plot shows that Toomre parameter increases due to the existence of winds or outflows, except for the cases with $l=0$ and $0.5$ which are unphysical solutions, as we discussed. But except to the very inner part of the disc, generally, Toomre parameter is still larger than one even in the presence of winds. However, we note that  similarity solutions are not valid at the regions very close to the inner or outer parts of the system. Larger the accretion rate at the outer edge of discs (i.e. larger $\dot{m}_0$), the disc becomes more massive and the size of the  gravitationally unstable  inner region increases due to the existence of wind/outflow.

\section{Discussion}

Our similarity solution show that the transition radius increases linearly with time, i.e. $r_{\rm tr} \approx \alpha' c_{\rm s} t$. Assuming that the radius $r_{\rm tr}$ is very close to the central part of the disc initially, there is not "no-wind" region initially and wind exists at all radii of the disc. At early times of evolution, wind does not modify the surface density significantly. As gas accretes toward the central parts, the surface density profile changes from inside. However,  the accretion rate is reduced  initially because of the wind, in comparison to the no-wind solution. Some fraction of the accreted mass and the angular momentum can be carried away to infinity by the wind, while leaving the remaining flow with smaller density at all parts and enhanced radial velocity at the outer part of the disc.

\begin{table*}\label{t}
 \centering
 \begin{minipage}{150mm}
\caption{Integrals $\mathcal{I}$ and $\mathcal{J}$ for the solutions with different input parameters. In the absence of winds/outflows, we have ${\mathcal I}=1.28$ and ${\mathcal J}=0$.}
\begin{tabular}{ c c c c   c c c c c c c c}
\hline\hline
&$(\Gamma_{0},l)=(0.1,1)$&&\vline&&$(\Gamma_{0},s)=(0.1,0.7)$&&\vline&&$(l,s)=(1.0,0.7)$&\\
\hline
$s$&${\mathcal I}$&${100 \mathcal J}$& \vline &$l$&${\mathcal I}$&${100\mathcal J}$&\vline&$\Gamma_{0}$&${\mathcal I}$&${100\mathcal J}$&\\
$0.1$&$0.83$&$7.19$&\vline&$0.0$&$1.84$&$9.81$&\vline&$0.01$&$0.88$&$4.49$\\
$0.5$&$0.87$&$5.08$&\vline&$1.0$&$0.88$&$4.49$&\vline&$0.05$&$1.04$&$2.66$\\
$0.7$&$0.88$&$4.49$&\vline&$1.5$&$0.58$&$2.86$&\vline&$0.1$&$1.22$&$0.06$\\
\hline\hline
\end{tabular}
\end{minipage}
\end{table*}

We can apply our similarity solutions to a self-gravitating disc just before forming a central star, namely, in the runaway collapse phase. This phase of evolution would correspond to the very early (class 0) stage.
A molecular cloud core can be approximated by an isothermal gas with sound speed $c_{\rm s} \sim 0.2 \rm Km$ s$^{-1}$
(e.g., Hayashi \& Nakano 1965). Typical age is considered to be $5\times 10^{4}$ yr. The radius $r=0.01$ pc
at $t_{0}=5\times 10^{4}$ yr corresponds to $x=1$ in our similarity variable. Then, the radius of the disc $r_{\rm d}$ becomes
\begin{equation}
r_{\rm d} = 0.01 {\rm pc} (\frac{x}{1}) (\frac{c_{\rm s}}{0.2 {\rm Km s^{-1}}})(\frac{t}{5\times 10^{4} {\rm yr}}).
\end{equation}

Now, we can calculate the mass of the disc as follows
\begin{equation}
M_{\rm disc}=\int_{0}^{r_{\rm d}} 2\pi \Sigma r dr,
\end{equation}
or
\begin{equation}
M_{\rm disc} = (9.51\times 10^{-2} M_{\odot})(\frac{t}{5\times 10^{4} {\rm yr}}) \mathcal{I},\label{eq:Mdisc}
\end{equation}
where ${\mathcal I} = \int_{0}^{1} x\sigma dx$. Having our similarity solutions, we can simply calculate this integral for different set of input parameters. When there is not a wind or outflow, the integral becomes ${\mathcal I} = 1.28$. But existence of a wind or outflow decreases the integral by a factor up to 2 depending on the input parameters (see Table 1). In other words, the mass of a self-gravitating disc with wind is reduced comparing to a similar disc but without wind. So, reduction to the mass of a disc with wind is mainly due to the decreased surface density at all  regions of the disc. This implies that the central mass object is forming up to two  times slower than a system without wind. Of course, this factor may increase if we consider cases, in which more angular momentum is carried away by wind (i.e., larger $l$).
We can determine how long is needed to increase the mass of the disc by one solar mass. According to equation (\ref{eq:Mdisc}), we can write
\begin{equation}
\tau \simeq \frac{0.52\times 10^6}{\mathcal{ I}} {\rm yr}.
\end{equation}
For a disc without wind, the mass of the disc increases by one solar mass within approximately  $\tau \simeq 4.0 \times 10^{5} {\rm yr}$. But the integral $\mathcal{ I}$ decreases by a factor up to $2$ according to Table 1. This allows the formation of a central core with one solar mass in $2$ times slower than a similar system without wind, i.e. $\tau \simeq 9 \times 10^{5} {\rm yr}$. This result is consistent with numerical simulations of early stages of massive discs with winds/outflows (e.g., Banerjee \& Pudritz 2007). Numerical study of Banerjee \& Pudritz (2007) is concentrated towards analyzing massive star formation and inseparable links between gravitational collapse and early wind-driven outflows. They showed that the disc, in the early stages of formation of a high mass star, is more massive than the protostar that is forming within it (see also Banerjee \& Pudritz 2006). Dominance of  mass of the disc is kept within $7\times 10^{4} {\rm yr}$ according to the simulations of Banerjee \& Pudritz (2006). In our paper, we neglected mass of the central protostar at early stages evolution which is a good approximation at early stages of formation. Also, the other analytical studies show that the central core grows to one solar mass in $1.6\times 10^{6}$ ${\rm yr}$ if other effects are ignored (e.g., Tsuribe 1999)

Although we presented the ratio of the total mass loss rate by wind to the accretion rate at each radius of the disc in Figures {\ref{fig:f1}}, {\ref{fig:f2}} and {\ref{fig:f3}}, it is desirable to  calculate how much mass is lost by wind during the early stages. Actually, observations of different systems show that the ratio of mass loss rate by wind to the accretion rate is around $0.1$ (e.g., Konigl \& Pudritz 2000). Having the self-similar solutions, one can simply show that the total mass-loss
rate by the wind is
\begin{equation}
{\dot M}_{\rm w}=(1.9\times 10^{-6} M_{\odot}/{\rm yr}) {\mathcal J},
\end{equation}
where ${\mathcal J}=\int_{0}^{1} x \sigma \Gamma dx$. Obviously, exact value of this integral depends on the profiles of the physical
 quantities of the disc, i.e. $\sigma$ and $\Gamma$. Table 1 shows this integral for the solutions with different sets of the input parameters. We can simply
 show that during  formation of a core with one solar mass the amount of mass loss by the wind is ${\mathcal J}/\mathcal{I}$ times the accreted mass. This ratio
 varies approximately from $0.001$ to $0.05$ according to Table 1. Thus, our mass loss rate is consistent with observations of protostellar systems.

Our solutions are valid as long as a  central mass has not been formed and the outer disc is not depleted with gas. Thus, there is no surprise  that the similarity solution demands a diverging total mass like a solution without wind, if $x$ tends to {\it infinity} (Mineshige \& Umemura 1997). But we restricted our solutions within $x\leq 1$, and so, the total mass of the disc is finite.  Once the outer part of the disc is depleted with gas and a central mass is formed, we can not apply our similarity solution with wind. However, the effects of wind on the early evolution of self-gravitating discs are remarkably worth to be considered, in particular reduced surface density and the accretion rate at all parts of the disc.

\section*{Acknowledgments}

I thank an anonymous referee for his/her constructive comments.

{}

\end{document}